\documentclass[twocolumn,showpacs,preprintnumbers,amsmath,amssymb]{revtex4}

\usepackage{graphicx}
\usepackage{dcolumn}
\usepackage{bm}

\newcommand{\beq}{\begin{equation}}
\newcommand{\eeq}{\end{equation}}
\newcommand{\beqn}{\begin{eqnarray}}
\newcommand{\eeqn}{\end{eqnarray}}
\newcommand{\be}{\begin{equation}}
\newcommand{\ee}{\end{equation}}
\newcommand{\bea}{\begin{eqnarray}}
\newcommand{\eea}{\end{eqnarray}}

\def\[{\left [}
\def\]{\right ]}
\def\({\left (}
\def\){\right )}

\def\r{\rho}

\def\m{\mu}

\def\r2{\sqrt{2}}

\def\del{\partial}
\def\sech{{\rm sech}}

\begin{document}

\preprint{HIP-2009-18/TH}

\title{Dark Solitons in Holographic Superfluids}

\author{V. Ker\"anen}
\email{Ville.Keranen@helsinki.fi}
 \altaffiliation[Also at ]{Department of Physics,
University of Helsinki, PO. Box 64, 00014
University of Helsinki, Finland.}
\author{E. Keski-Vakkuri}%
 \email{Esko.Keski-Vakkuri@helsinki.fi}
\author{S. Nowling}
\email{Sean.Nowling@helsinki.fi}
\altaffiliation[Also at ]{
Department of Mathematics and Statistics,
University of Helsinki,
PO. Box 64, 00014 University of Helsinki, Finland.}
\author{K. P. Yogendran}
\email{yogendran.kalpat@helsinki.fi}
\affiliation{
Helsinki Institute of Physics,\\
P.O. Box 64, 00014 University of Helsinki, Finland
}

\date{\today}

\begin{abstract}
We construct dark soliton solutions in a holographic model of a
relativistic superfluid. We study the length scales associated with the
condensate and the charge density depletion, and find that the two scales differ
by a non-trivial function of the chemical potential. By adjusting the chemical potential,
we study the variation of the depletion of charge density at the
interface.
\end{abstract}

\maketitle

\paragraph{\label{intro}Introduction.}
One of the exciting theoretical developments in recent years has been
the evolution of holographic gauge/gravity duality.  The holographic
principle, a dimensional reduction in the reorganization of all
information, was proposed to be a general feature of any gravitational
theory \cite{holo}.  So far the most concrete realization is the
correspondence between ${\cal N }=4$ supersymmetric $SU(N)$ Yang-Mills
gauge theory in 3+1 dimensions and type IIB supersymmetric string
theory in 9+1 dimensions \cite{Maldacena:1997re}, where the spacetime
manifold consists of a 4+1 dimensional anti-de Sitter spacetime and a
5-dimensional sphere, with the structure $AdS_5\times S^5$. The
mapping of gauge theory into string theory becomes tractable and
useful in the strong t'Hooft coupling $\lambda = g^2_{YM}N$ limit of
the Yang-Mills theory, which corresponds to the low energy
supergravity limit of string theory in weakly curved spacetime. The
gravitational theory can then be Kaluza-Klein reduced to the $AdS_5$
spacetime, yielding the holographic gauge-gravity duality between
strongly coupled gauge theory in 3+1 Minkowski space and gravitational
theory in 4+1 dimensions. There are many ways to deform and extend the
theory to obtain other cases of dual theories, some of which are
reviewed in \cite{Aharony:1999ti}.

The duality is not only a profound theoretical statement about a
non-gravitational theory secretly being a theory of gravity in higher
dimensions (or vice versa). It is also a useful calculational tool, in
particular for computing correlation functions in a strongly coupled
gauge theory, when perturbative methods are no longer tractable
\cite{corre}. Most recently this connection has been applied to
various strongly interacting condensed matter systems (reviewed {\em
  e.g.} in \cite{Hartnoll:2009sz,Sachdev:2008ba,Herzog:2009xv}). There
are various constructions for holographic gravity duals, allowing new
insights and methods for the analysis of correlation functions.  In
this document we investigate a recently formulated gravitational dual
theory \cite{Hartnoll:2008vx,Gubser:2008px}, see also
\cite{Herzog:2008he,Hartnoll:2008kx,Basu:2008st}, in the context of
(relativistic) superfluids with a spontaneously broken global U(1)
symmetry.  We will work in the so called probe limit where we can neglect gravitational backreaction to the metric.  This backreaction would be important as one approaches zero temperature.

A characteristic structure supported by superfluids is a
vortex. However, in this work, we will investigate other solitonic
configurations which interpolate between two phases.  Vortex solutions
were recently reported in the holographic model \cite{Hartnoll:2008vx}
by different groups
\cite{Albash:2009ix,Albash:2009iq,Montull:2009fe}, in the
context of interpreting \cite{Hartnoll:2008vx,Hartnoll:2008kx} as a
dual model of a Type II superconductor.

As one source of inspiration, we note the fascinating crossover from
BCS superconductors to BEC superfluids which can be experimentally
realized in ultracold gases of fermionic atoms.  Secondly we note that
the condensate superfluid can support interesting localized but
extended defects, called bright or dark solitons -- interfaces of
increased or reduced charge density between two superfluid phases (for
a review, see {\em e.g.} \cite{kivshar1998, kevrekidisPG2008}).  They
can be created in experiments, and have many interesting and
incompletely understood properties. In particular one can study what
happens to a dark soliton during the BCS-BEC crossover. In a
theoretical study it was found that during the transition to the BCS
regime, the charge depletion diminishes and becomes visibly modulated
by Friedel oscillations \cite{antezzaetal}.  Finally, we note that at
strong coupling dark solitons, in particular multidimensional ones,
are difficult to study with standard theoretical tools and {\em e.g.}
the spectrum of excitations is ill-understood.

In this document we construct dark soliton solutions at finite
temperature in (relativistic) holographic superfluids, with a dual
gravitational description that describes strongly coupled dynamics in
field theory. We find that the depletion of charge density varies as
we dial the only available parameter of the model, the chemical
potential.  The existence of such solutions might have been anticipated due to generic topological arguments, but their detailed properties could very well be different than those of the solution to the GP equation.  Indeed, we find that there are qualitative differences.  The depletion fraction is far from 100\% unlike the GP example.  Also there are two characteristic length scales governing the condensate density and charge density as opposed to a single length scale in the GP equation.

This document is organized as follows. We begin with a brief
discussion of the dark soliton and its properties using the
Gross-Pitaevskii equation as a guideline. This will enable us to
determine quantities of interest which will be computed using the
holographic description. The configurations of interest are obtained
by solving a system of partial differential equations in AdS space. We
present the equations and the boundary conditions which sustain the
soliton. We then discuss the numerical methods and present the
results. We conclude with a brief discussion of the results and future
directions.

\paragraph{Dark Solitons in Superfluids.}
Dark solitons are spatial interfaces of reduced charge density between two
superfluid phases. As a starting point, we can model them by using the
time independent Gross-Pitaevskii (GP) equation for the superfluid order
parameter $\Phi$
\be
-\frac{1}{2m_B}\del_x ^2\Phi+(2V-\m)\Phi+ g\Phi |\Phi|^2=0 \label{GPeqns}
\ee
where the coefficients depend on temperature and
chemical potential.
The dark soliton is a spatially varying solution
which interpolates between the potential minima (phases)
\be
\Phi(\infty)=\Delta,\, \Phi(-\infty)= -\Delta; \qquad
\Delta=\sqrt{\frac{(\m-2V)}{2g}}.
\ee
For fermionic systems, this equation may be obtained from the
microscopic Bogoliubov-de Gennes equation \cite{PieriStrinati} (in a
strong coupling limit) as a mean field description - which
nevertheless may be expected to give a reasonably accurate description
of the essential physics. However, we will not assume any particular
dependence for the coefficients.

The GP equation (\ref{GPeqns}) has a well known exact soliton solution
\beqn
\Phi=\sqrt{\frac{\mu-2V}{2g}}\tanh(x/\xi),
\quad \rho(x)= q |\Phi|^2\label{eq:GPdark}
\eeqn
which interpolates between the two vacua $\Phi=\pm\Delta$.
Here $\rho$ is the charge density and $q$ is the unit of U(1) charge.
The coherence length $\xi$, can then be written in a useful form in terms
of the parameters of the equation
\beqn
\xi=\frac{1}{\sqrt{4gm_B}\Delta} \quad =\frac{1}{\sqrt{2(\mu-2V)m_B}}.
\label{eq:GL5}
\eeqn
This will act as a guide in determining the quantities of interest for
the holographic bulk description without assuming any particular
dependence for the coefficients (or even assuming the GP equation).

\paragraph{Holographic Description. \label{HoloEqns}}
The holographic modelling of the superfluid system was constructed in
\cite{Hartnoll:2008vx} following the ideas in \cite{Gubser:2008px}. We
consider a Maxwell-Scalar system in a 4-D planar AdS black hole
background with a metric
\be
ds^2 = L^2(-\frac{f dt^2}{z^2}+\frac{dz^2}{f z^2}+\frac{ d\vec{x}^2}{z^2})
;\,\, f=1-\left(\frac{z}{z_T}\right)^3
\ee
where $L$ is the AdS radius and which will be set to unity in the following
discussion. The bulk Lagrangian is taken to be
\be
{\mathcal L}=\sqrt{-g}(-\frac{1}{4} F_{\mu\nu} F^{\mu\nu}
-D_\mu\Psi D^\mu \bar\Psi+2\Psi\bar\Psi).
\ee
In the gauge $A_z=0$, the nontrivial equations of motion for the bulk
fields $\Psi=\frac{1}{\sqrt{2}}z\tilde R$ and $A_0=A$, after suitable
rescaling, are
\beqn\label{AdSeqs}
f\tilde R''+f' \tilde R'-z \tilde R +\partial_x ^2 \tilde R +
\tilde R (\frac{A^2}{f})=0\\
f A''+\partial_x ^2 A-\tilde R ^2 A=0.\nonumber 
\eeqn

Using the equations, it can be easily seen that, close to the boundary
at $z=0$, the fields $\Psi, A$ behave as
\be
\tilde R =  \tilde R^{(1)}+z \tilde R^{(2)}+... \quad A = A
^{(0)}+z A ^{(1)}+...
\ee
in an expansion along the $z$-direction.

The dual field theory (in this case, the superfluid system) is said to
be at a temperature $T=T_H=\frac{3}{4\pi z_T}$ which is the Hawking
temperature of the black hole in AdS space.  According to the recipe
provided by the AdS/CFT correspondence, the boundary values of the
bulk fields are related to the (dual) superfluid system as follows
\begin{eqnarray}
\tilde R^{(1)}= \sqrt{2}z_T J(x)&,&\qquad
\tilde R^{(2)}= \sqrt{2} z_T ^2\, \langle O_2\rangle(x)\\
A^{(0)}= z_T\, \mu(x)&,& \qquad
A^{(1)} = z_T ^2\,\rho(x)
\label{definitions}
\end{eqnarray}
where $\mu$ is the chemical potential, $\rho$ is the (total) charge
density, $O_2$ is a charged operator of mass dimension 2 and $J(x)$ is
to be regarded as an external source in the field theory for this
operator. We will set such sources to zero in this work. The
expectation values of this operator (in the ground state) may be taken
to be the order parameter of the superfluid system (since
superfluidity is the spontaneous breaking of a global U(1) symmetry).  We will work in the probe limit of this background where one may ignore gravitational backreaction.  Such backreation would become important near zero temperature.

The solution is then uniquely determined by the boundary conditions
\be
\tilde R^{(1)}= 0, \quad
A^{(0)}= z_T \mu={\rm constant} \quad
\ee
and regularity conditions at the horizon of the black hole (one each
for $R$ and $A_0$).

In this particular system, it is also possible to identify $\tilde
R^{(1)}$ with the vev of a charged operator of mass dimension $1$
in which case $\tilde{R}^{(2)}$ has the source interpretation in the
dual theory. This case will be discussed in a later publication.

In the work \cite{Hartnoll:2008vx}, it was shown that for sufficiently
small $T$ one can find solutions where the scalar field is
nonzero. That is to say, for low enough temperature we can have
nonzero charged condensates,and hence superfluidity.  Since there is
only one independent parameter determining the solution of
(\ref{AdSeqs}) we can equivalently vary the chemical potential $\mu$
and find the condensate phase at large enough $\mu$.  The critical
value of $\mu$ depends on whether we consider $\tilde{R}^{(1)}$ or
$\tilde{R}^{(2)}$ as the order parameter.

\paragraph{Numerical method. \label{Numerics}}
We place the system in a large box, $(z,x)\in [0,1]\times
[-L_x/2,L_x/2]$ and use numerical methods to solve the field equations
(\ref{AdSeqs}). Searching for dark soliton solutions, the basic
strategy is to assume an initial configuration with an interface at
$x=0$ which satisfies the boundary conditions and then use the
Gauss-Seidel relaxation method to obtain the solution.  The topological information is included with the initial seed configuration and is preserved during relaxation.  We have used Neumann boundary conditions at $x=\pm L$ and confirmed that the results are insensitive to the size $L$.  For derivatives in Gauss-Seidel, we use a second order representation in the bulk and a one sided, third order representation at $x=\pm L$.

In the iteration method there is a danger that what looks like a
solution ceases to be one after suitably many iterations. However,
consider the topological stability criterion in two dimensions, the
space dimension of $\partial $AdS$_4$ which is the space in which the
dual theory resides. The spacetime independent solutions to
(\ref{GPeqns}) define a space of vacua $\mathcal{M}$. In our case
clearly $\mathcal{M}= U(1)$. Topologically stable one-dimensional
objects (kinks) are not possible in this system because $\mathcal{M}$
is arc-connected. But topologically stable zero-dimensional objects
(vortices) are possible since $\pi_1(\mathcal{M})=\mathbb{Z}$.

However, the starting point for dark soliton solutions is (\ref{AdSeqs})
which involves the {\em real} valued field $\tilde{R}$ and have a
$\tilde{R}\mapsto -\tilde{R}$ symmetry. So the $U(1)$ symmetry has
been reduced to $\mathbb{Z}_2$. Moreover, we require translational
invariance in the $y$ direction which means that the spatial boundary
is also $\mathbb{Z}_2$. Thus, the solutions we discuss are
topologically {\em nontrivial} maps from $x=\pm\infty$ to
$\mathbb{Z}_2$.

This provides the underlying reason that we expect our numerical
solutions to be stable approximations to the actual solutions.
Further, we find that the solutions we obtain asymptote to those of
\cite{Herzog:2008he} far away from the interface. We also
find that even if we perturb these configurations (preserving the
boundary conditions), the perturbations decay rapidly (in iteration
time).  

We have identified two main sources of error: discretization error from the finiteness of the lattice and algebraic error from solving the discretized equations at each site.  By substituting the numerically determined solutions into the differential equations we have checked that the error decreases monotonically as one increases the lattice size. 

The dark soliton is known to be an unstable object in two or more
dimensions via the snake instability. This happens as follows. If we
break the translational invariance in $y$ direction (or
reinstate the phase d.o.f.) by small perturbations, the superfluid
dark soliton will decay into topologically stable vortices by snake
instability (see for example \cite{Federetal} and references therein),
which has been conjectured as due to tachyonic states in the
Bogoliubov spectrum around the dark soliton.

\begin{figure}[h!]
\includegraphics[scale=0.4]{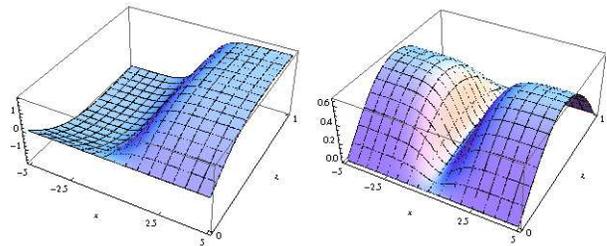}
\caption{\label{fig1}Profiles for the scalar field $\tilde R$ (left) and the
  normalized gauge field $A-z_T\mu(1-z)$ (right) in the bulk for $\mu=5.18$.}
\end{figure}

\begin{figure}[h!]
\includegraphics[scale=0.65]{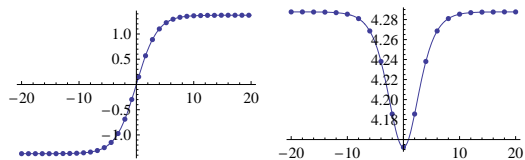}
\caption{\label{fig:FITS} Condensate and charge density vs. $x$.}
\end{figure}

\paragraph{Results.}
Generically, the solutions for the scalar field $\tilde R$ and the
vector field $A$ in the bulk look as in Figure \ref{fig1}.  Note that
all graphs are plotted using dimensionless variables (denoted with a
tilde) - dimensions maybe restored by using eqn(\ref{definitions}).
In order to highlight the structure of the solution for the vector
field, we have subtracted a ``background'' contribution and plot
$A-z_T\mu(1-z)$ instead of $A$. From the numerical solutions we
obtain, we can plot the boundary profiles of the charge density
$A^{(1)}$ and the condensate $\tilde{R}^{(2)}=(\partial_z\tilde
R)|_{z=0}$. The numerical data and expectations from the GP
equations suggest that the condensate can be fitted by a
$\tanh(\frac{x}{\xi})$ profile and the charge density by a
$\sech^2(\frac{x}{\xi_q})$ profile (with {\em a priori} unrelated
coherence  lengths to be determined from the fit). The data along
with the fitted curves are shown in the Figure \ref{fig:FITS}.
Using a least square fit, we then extract the coherence length $\xi$
from the condensate profile and $\xi_q$ from the charge density.

\begin{figure}[h!]
\includegraphics[scale=0.5]{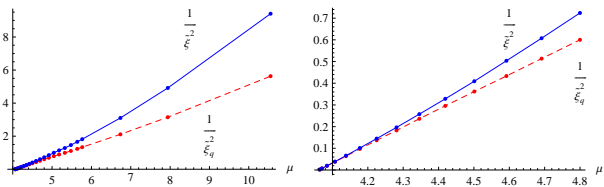}
\caption{\label{Figximu} [Coherence length]$^{-2}$ vs. chemical potential.}
\end{figure}

\begin{figure}[h]
\includegraphics[scale=.5]{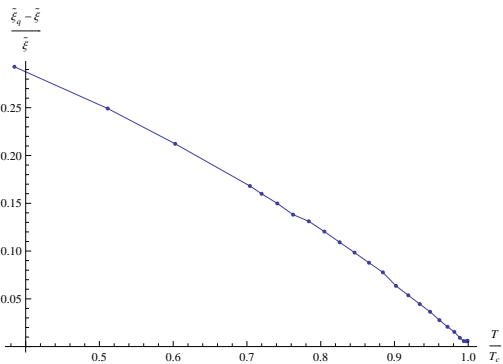}
\caption{\label{Fig:diff}
Difference in Length scale vs. Temperature.}
\end{figure}

Motivated by the earlier discussion of the GP equations, we plot
the dependence of the coherence lengths on the chemical potential $\mu$ in Figure \ref{Figximu}.  Using the degree to which our solutions satisfy the differential equations as an estimate of the total error, we find that the error bars on the curves are too small to be visible.
One might anticipate a divergence as we reach the
critical value for the chemical potential. We find that for small
values of the chemical potential
\beqn
\tilde\xi(\mu)\approx 0.99{(\tilde\mu-\tilde\mu_{0})^{-\frac{1}{2}}},
\label{eq:DS1}
\eeqn where $\tilde \mu=A(z=0)=\mu z_T$ is the dimensionless chemical
potential and $\tilde\mu_{0}\approx 4.07$ is the critical value below
which there is no condensate for $\langle\mathcal{O}_{2}\rangle$
(cf. eqn. (\ref{eq:GL5})).  As can be seen from equation
(\ref{eq:GPdark}) the two scales $\xi$ and $\xi_q$ are the same in GP
model. However, in the holographic model, it is seen that the two
scales are different and their difference is a non-trivial function of
the temperature (or inverse chemical potential) as displayed in figure
\ref{Fig:diff}.
As might be expected, the difference between the two scales tends to
zero as we approach the critical chemical potential (or temperature).
However, for low temperatures (or large chemical potentials), the
difference appears to asymptote to a constant value. This is a very
exciting feature because the presence of two length scales would be detectable in any physical realization of the dual field theory.

For the solutions discussed in
\cite{Herzog:2008he}, we know the dependence of the condensate
$\langle\mathcal{O}_{2}\rangle$ on the chemical potential (for values
close to the critical value $\tilde\mu_0$)
\beqn
z_T ^2\langle\mathcal{O}_{2}\rangle\approx 3.46\sqrt{\tilde\mu-\tilde\mu_{0}}.
\label{eq:DS2}
\eeqn
Using equations (\ref{eq:DS1}) and (\ref{eq:DS2}) we get the
connection between the coherence length and the condensate for small
chemical potentials
\beqn
z_T ^2 \tilde\xi\approx 0.29 /\langle\mathcal{O}_{2}\rangle \ .\label{eq:DS3}
\eeqn

\begin{figure}[h]
\includegraphics[scale=.6]{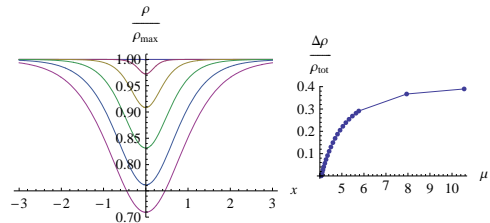}
\caption{\label{Fig:dep-profile}
Density depletion profile vs. chemical potential.}
\end{figure}

Another quantity that is of great interest is the amount of depletion
of charge density at the interface and its variation with the chemical
potential for fixed temperature. This is shown in the Figure
\ref{Fig:dep-profile} where we plot the spatial variation of the
charge density for a few values of the chemical potential. A striking
feature is that the charge density does not appear to vanish at the
interface even at zero temperature in sharp contrast to the dark
soliton solution of the GP equation. The (total) density depletion
fraction increases monotonically with increasing chemical potential.
In the second graph in Figure \ref{Fig:dep-profile}, we see
that the percentage of depletion seems to tend to a constant as we
tend to zero temperature (or large chemical potential).  However, our numerical results are less accurate for the larger values of the chemical potential since it is
computationally much more expensive to obtain solutions when the field
values become large.

Note that our system is at finite temperature and therefore we might
expect that only a part of the total charge density comes from the
condensate, while some of it is due to quasiparticles which are not in
the condensate. This is supported by the fact that the condensate
vanishes at a nonzero value of the charge density.



\paragraph{Discussion.}
In this work we have shown that the holographic model of a superfluid
constructed in \cite{Hartnoll:2008vx} supports dark solitons, that
are stationary structures in the condensate, stabilized by non-linear
interactions. At the simplest level such solutions can be obtained from
the Gross-Pitaevskii equation (\ref{GPeqns}), which includes the necessary
interaction terms to stabilize the dark soliton. Still the GP equation
has its limitations since it is applicable only in the limit of small
temperatures and densities.

Our work shows that the functional form of the condensate profile
of the holographic dark soliton is similar to the one obtained
from the GP equation, but there are striking differences. First of all the charge density does
not go to zero at the core of the dark soliton, which means
that not all of the particles are in the condensate. The density
depletion is seen to vary as a function of the chemical potential.

Furthermore we find that the length scale associated with the
charge density depletion is different from the coherence length
determined from the condensate. This difference is observed to
be a non-trivial function of the chemical potential. It would
be interesting to see if such a difference of scales of the
dark soliton could be seen in currently available experimental systems.

It will be important to study dark solitons in non-relativistic duals that are closer to BEC-BCS systems, when they become available. However, we find it interesting that even in a relativistic dual, dark solitons have similar features as those observed in BEC-BCS systems.


In a longer forthcoming manuscript, we will study dark solitons in the
$\langle\mathcal{O}_1\rangle$ theory, which turn out to have
interesting differences to the ones studied here.  In the immediate
future, one might also construct vortices and compare them with the
dark solitons \cite{Randeria}.  In a closely related work, the
hydrodynamic properties of this system was studied \cite{Kaminski}. It
would be important to examine the relation between that study and
ours.

It will of course be interesting to study the effects of magnetic
fields since one can use an external magnetic field to control the
density of states of this system. Another exciting prospect is the
Abrikosov vortex lattice. An important shortcoming of our work is that
we work in the probe limit. Taking the gravitational backreaction into
account would be doubly interesting - both from the field theory point
of view as well as from the gravitational point of view. Some of these
projects are currently under study.

{\em Acknowledgments} We thank C.~V.~Johnson and R.~G.~Leigh for
comments, and in particular L.~D.~Carr and S.~Vishveshwara for helpful
comments on dark solitons.  We would also like to thank Ari Harju for
a very useful discussion which sparked off this study.  The work of
V.K. and E.K-V. has been supported in part by the Academy of Finland
grant nr 1127482. E.K-V. and S.N. thank the Aspen Center for Physics
for hospitality and for providing an opportunity for illuminating
discussions while this work was in progress.

\end{document}